\documentclass{emulateapj}
\usepackage{apjfonts}

\newcommand{\f}{\phantom{2}}
\newcommand{\mc}{\multicolumn}

\newcommand{\gtsimeq}{\raisebox{-0.6ex}{$\,\stackrel 
        {\raisebox{-.2ex}{$\textstyle >$}}{\sim}\,$}}

\newcommand{\mgii}{Mg\,{\sc ii}}

\newcommand{\lya}{Ly\,$\alpha$}
\newcommand{\lyb}{Ly\,$\beta$}
\newcommand{\lyg}{Ly\,$\gamma$}

\newcommand{\myemail}{chris.willott@nrc.ca}

\def\xmm{{\it XMM-Newton~}}

\def\co21{CO\,(2-1)}

\shorttitle{Six quasars at redshift 6 discovered by the CFHQS}
\shortauthors{Willott et al.}

\begin{document}


\title{Six more quasars at redshift 6 discovered by the Canada-France High-$z$ Quasar Survey}


\author{
Chris J. Willott\altaffilmark{1},
Philippe Delorme\altaffilmark{2},
C\'eline Reyl\'e\altaffilmark{3},
Loic Albert\altaffilmark{4},
Jacqueline Bergeron\altaffilmark{5},
David Crampton\altaffilmark{1},
Xavier Delfosse\altaffilmark{2},
Thierry Forveille\altaffilmark{2},
John B. Hutchings\altaffilmark{1},
Ross J. McLure\altaffilmark{6},
Alain Omont\altaffilmark{5},
and David Schade\altaffilmark{1},
}

\altaffiltext{1}{Herzberg Institute of Astrophysics, National Research Council, 5071 West Saanich Rd, Victoria, BC V9E 2E7, Canada; \myemail}
\altaffiltext{2}{Laboratoire d'Astrophysique, Observatoire de Grenoble, Universit\'e J. Fourier, BP 53, F-38041 Grenoble, Cedex 9, France}
\altaffiltext{3}{Institut Utinam, Observatoire de Besan\c{c}on, Universit\'e de Franche-Comt\'e, BP1615, 25010 Besan\c{c}on Cedex, France}
\altaffiltext{4}{Canada-France-Hawaii Telescope Corporation, 65-1238 Mamalahoa Highway, Kamuela, HI96743, USA}
\altaffiltext{5}{Institut d'Astrophysique de Paris, CNRS and Universit\'e Pierre et Marie Curie, 98bis Boulevard Arago, F-75014, Paris, France}
\altaffiltext{6}{Scottish Universities Physics Alliance, Institute for Astronomy, University of Edinburgh, Royal Observatory, Blackford Hill, Edinburgh, EH9 3HJ, UK}

\begin{abstract}

We present imaging and spectroscopic observations for six quasars at
$z \ge 5.9$ discovered by the Canada-France High-$z$ Quasar Survey
(CFHQS). The CFHQS contains sub-surveys with a range of flux and area
combinations to sample a wide range of quasar luminosities at $z\sim
6$. The new quasars have luminosities 10 to 75 times lower than the
most luminous SDSS quasars at this redshift. The least luminous
quasar, CFHQS\,J0216-0455 at $z=6.01$, has absolute magnitude
$M_{1450}=-22.21$, well below the likely break in the luminosity
function. This quasar is not detected in a deep \xmm survey showing
that optical selection is still a very efficient tool for finding
high redshift quasars.

\end{abstract}


\section{Introduction}

Observations of distant galaxies and quasars give us a direct view of
the past because of the finite speed of light. We can now discover
these objects at redshifts $z>6$, revealing the nature of the
universe during the first billion years. 

Quasars are particularly useful because their high intrinsic
ultraviolet luminosities can be used as a background light source to
be absorbed by intervening gas. Studies of quasars out to redshifts of
$z=6.4$ have put the strongest constraints yet on when the universe
was reionized (Fan et al. 2006a). The absence of Gunn-Petersen troughs
along most lines of sight at $z<6$ shows that cosmic reionization was
just about complete by $z=6$. The high electron scattering optical
depth of microwave background photons shows that the first phase of
reionization occurred significantly earlier than $z=6$ (Dunkley et
al. 2008). The reionization history at $z>6$ is still quite uncertain.
Spectroscopy of quasars at redshifts between $z=6$ and $z=6.5$ provides
the most detailed information currently available on this period
(White et al. 2003; Wyithe et al. 2008).

It is now widely believed that the high luminosities of quasars are
generated via gravitational potential energy released as matter falls
towards supermassive black holes. Reverberation mapping has shown the
existence of a relationship between broad emission line velocity,
luminosity and black hole mass which allows black hole masses to be
estimated for all quasars (Kaspi et al. 2005; Peterson et
al. 2004). Studies of the highest redshift quasars therefore allow
inferences on the growth of the most massive black holes in the early
universe (Willott et al. 2003; Jiang et al. 2007; Kurk et al. 2007).
The black hole masses implied lead to strong constraints on theories
for the early growth of supermassive black holes in galaxies
(e.g. Volonteri \& Rees 2005; Li et al. 2007).

The Canada-France High-$z$ Quasar Survey (CFHQS) is a survey based on
multi-color optical imaging at the Canada-France-Hawaii Telescope
(CFHT) to discover quasars at redshifts $5.8<z<6.5$ much less luminous
than those discovered by the Sloan Digital Sky Survey (SDSS). The
CFHQS quasars are 10 to 75 times less luminous than the main SDSS
sample (Fan et al. 2006b) and 5 to 25 times less luminous than the SDSS
deep stripe sample (Jiang et al. 2008). The range in luminosity is essential
for an accurate determination of the quasar luminosity function and
for studying lower mass black holes than those powering SDSS
quasars. We have previously reported the discovery of four quasars at
$z>6$ by the CFHQS (Willott et al. 2007; hereafter W07). In this
paper, we present the discovery of a further six quasars bringing the
total of CFHQS quasars at $z \ge 5.9$ to ten.

All optical and near-IR magnitudes in this paper are on the AB
system. Cosmological parameters of $H_0=70~ {\rm km~s^{-1}~Mpc^{-1}}$,
$\Omega_{\mathrm M}=0.28$ and $\Omega_\Lambda=0.72$ (Komatsu et
al. 2008) are assumed throughout.


\begin{figure}
\hspace{-0.2cm}
\resizebox{0.48\textwidth}{!}{\includegraphics{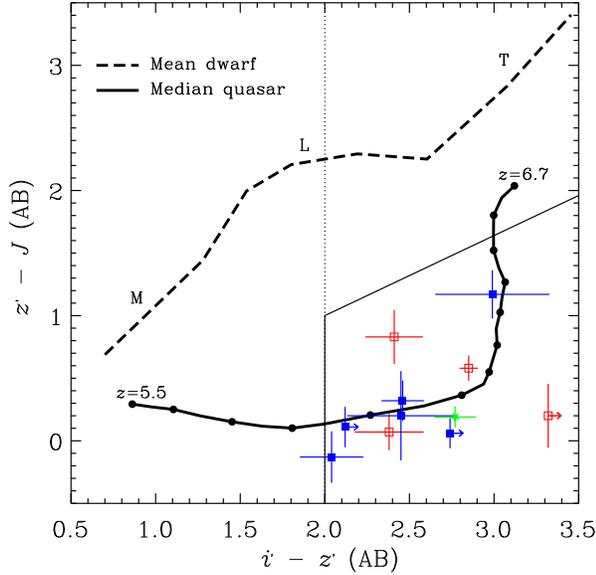}}
\caption{Color-color diagram showing the criteria used to select
quasar candidates (lower-right corner bounded by solid line). The
dotted line at $i'-z'=2$ shows the initial selection cut from
optical data. The thick solid line shows the expected color of a
template quasar redshifted from $z=5.5$ to $z=6.7$. The thick dashed
line shows the mean colors of a sample of M, L and T dwarfs. See
Willott et al. (2005), W07 and Delorme et al (2008) for details of
the expected quasar and dwarf colors. The six new CFHQS quasars are
shown with blue filled squares. SDSS\,J2315-0023 discovered by Jiang
et al. (2008) is shown as a green star. The four previous CFHQS
quasars from W07 are shown as red open squares. The many brown
dwarfs discovered by the CFBDS are not plotted here (see Delorme et
al. 2008).
\label{fig:colcol}
}
\end{figure}

\section{Observations}

\subsection{Imaging}

The six quasars whose discovery we present in this paper were found
from several different optical imaging surveys which all form part of
the CFHQS. The CFHQS uses several datasets with different magnitude
limits to sample a range of quasar luminosities. Because ultracool
brown dwarfs are also found from this type of optical imaging, the
CFHQS shares imaging data with the Canada-France Brown Dwarf Survey
(CFBDS). Further details on the imaging and data processing can be
found in W07 and Delorme et al. (2008). A plot showing the sky
locations of all the fields observed to the end of 2007 can be found
in Delorme et al. (2008). We give here a brief description of each
dataset.

It is important to note that CFHQS/CFBDS followup is still ongoing and
there are likely many more quasars to be found in this survey over the
next couple of years. For this reason, we defer discussion of the
$z=6$ quasar luminosity function to a future publication when there
will be some areas of the CFHQS with well defined selection criteria
and completeness.

\subsubsection{Red-sequence Cluster Survey 2 (RCS-2)}

The majority of the sky area in the CFHQS ($\sim 550$ square degrees)
is part of the RCS-2 survey (Yee et al. 2007). These data consist of
MegaCam $g'r'i'z'$ imaging with exposure times in each filter of
240\,s, 480\,s, 500\,s, 360\,s, respectively. The RCS-2 data contain
four of the quasars presented in this paper, CFHQS\,J0055+0146,
CFHQS\,J0102-0218, CFHQS\,J2318-0246 and CFHQS\,J2329-0403 and three
of the quasars presented in W07.

\begin{figure}
\hspace{-0.2cm}
\resizebox{0.48\textwidth}{!}{\includegraphics{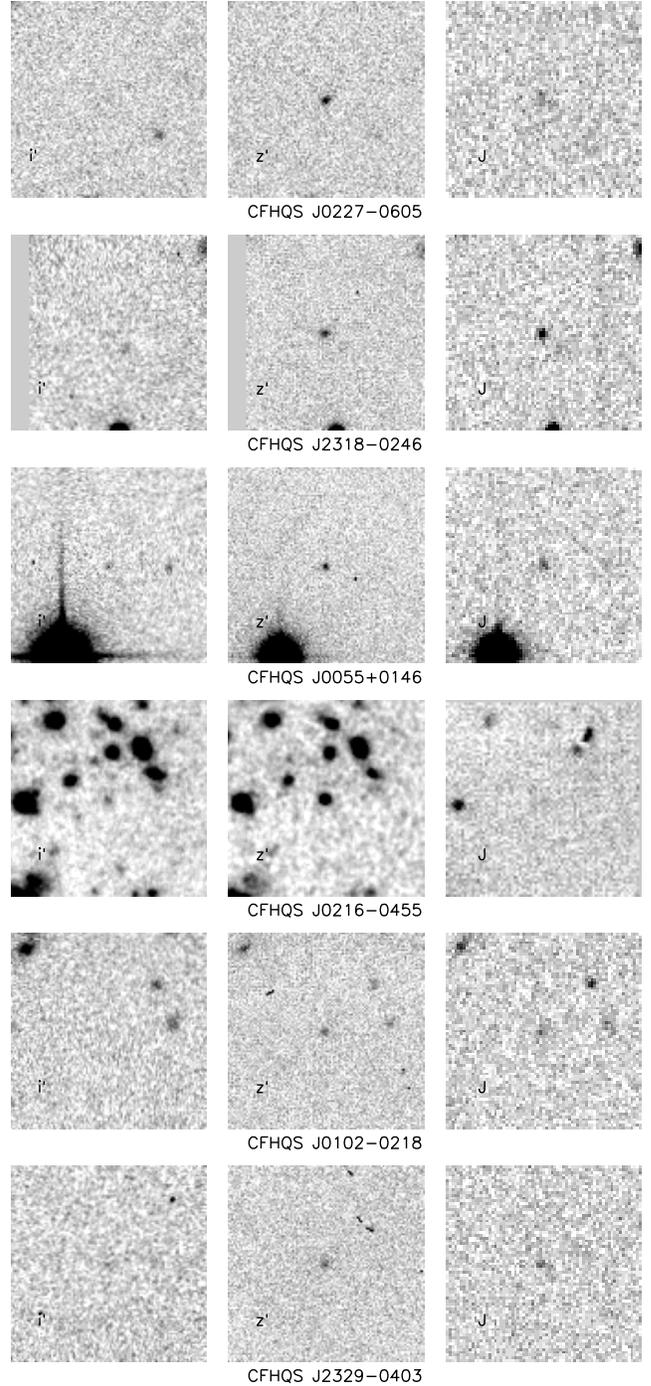}}
\caption{Images in the $i'$, $z'$ and $J$ filters centered on the six
  CFHQS quasars. Each image covers $20'' \times 20''$. The images are
  oriented with north up and east to the left.
\label{fig:cutouts}
}
\end{figure}

\subsubsection{CFHTLS Very Wide}

The CFHT Legacy Survey (CFHTLS) Very Wide is the shallowest component of the CFHTLS and
covers several hundred square degrees. The CFHQS contains $\sim 150$
square degrees of data from the Very Wide. The total exposure times
for the Very Wide are comparable to RCS-2 with 540\,s at $i'$ and
420\,s at $z'$. To date only one high redshift quasar has been
discovered in the CFHTLS Very Wide, CFHQS\,J1509-1749 from W07.

\subsubsection{CFHTLS Wide}

The CFHTLS Wide is the intermediate depth/area component of the
CFHTLS. It consists of 171 square degrees at $u^*g'r'i'z'$ with
typical MegaCam integration times of 4\,300\,s at $i'$ and 3\,600\,s
at $z'$. CFHQS\,J0227-0605, presented in this paper, is the first high
redshift quasar from the CFHTLS Wide.

\subsubsection{CFHTLS Deep}

The CFHTLS Deep consists of four MegaCam pointings each of $\sim 1$
square degree at $u^*g'r'i'z'$. Typical final integration times are
250\,000\,s at $i'$ and 200\,000\,s at $z'$.  The negative results of
a search for high redshift quasars in the first release of the Deep
(T0001 with $\approx 20$\% of the final integration time) was presented
by Willott et al. (2005). No high redshift quasars have yet been
discovered in the CFHTLS Deep.

\begin{table*}
\begin{center}
\caption{\label{tab:photom} Positions and photometry for the new CFHQS quasars} 
\begin{tabular}{clccccc}
\hline
\hline
\mc{1}{c}{Quasar} &\mc{1}{c}{RA and DEC (J2000.0)} &\mc{1}{c}{$i'$ mag} &\mc{1}{c}{$z'$ mag}&\mc{1}{c}{$J$ mag} &\mc{1}{c}{$i'-z'$} &\mc{1}{c}{$z'-J$} \\
\hline
CFHQS\,J005502+014618 &  00:55:02.91 +01:46:18.3 &  $24.65 \pm 0.11$           & $ 22.19 \pm 0.06$  & $21.87 \pm 0.15$ & $ 2.46 \pm 0.13$ & $0.32 \pm 0.17$ \\
CFHQS\,J010250-021809 &  01:02:50.64 -02:18:09.9 &  $>24.42$\tablenotemark{a}  & $ 22.30 \pm 0.08$  & $22.07 \pm 0.17$ & $ >2.12$         & $0.23 \pm 0.19$ \\
CFHQS\,J021627-045534 &  02:16:27.81 -04:55:34.1 &  $26.85 \pm 0.23$           & $ 24.40 \pm 0.06$  & $24.20 \pm 0.35$ & $ 2.45 \pm 0.24$ &$0.20 \pm 0.35$ \\ 
CFHQS\,J022743-060530 &  02:27:43.29 -06:05:30.2 &  $25.70 \pm 0.31$           & $ 22.71 \pm 0.06$  & $21.46 \pm 0.16$ & $ 2.99 \pm 0.32$ & $1.25 \pm 0.17$ \\ 
CFHQS\,J231802-024634 &  23:18:02.80 -02:46:34.0 &  $>24.40$\tablenotemark{a}  & $ 21.66 \pm 0.05$  & $21.60 \pm 0.11$ & $ >2.74$         & $ 0.06\pm 0.12$ \\ 
CFHQS\,J232914-040324 &  23:29:14.46 -04:03:24.1 &  $23.91 \pm 0.17$           & $ 21.87 \pm 0.08$  & $22.00 \pm 0.19$ & $ 2.04 \pm 0.19$ & $ -0.13\pm 0.15$ \\ 
\hline
\end{tabular}
\tablenotetext{ }{All magnitudes are on the AB system.}
\tablenotetext{a}{Where not detected at $> 2\sigma$ significance, a $2\sigma$ lower limit is given.}
\end{center}
\end{table*}

\begin{table*}
\begin{center}
{\caption{\label{tab:specobs} Optical spectroscopy observations of the new CFHQS quasars}}
\begin{tabular}{cccccccc}
\hline
\hline
\mc{1}{c}{Quasar} &\mc{1}{c}{Redshift} &\mc{1}{c}{Date~~~~~~~}&\mc{1}{c}{Resolving} &\mc{1}{c}{Slit Width} &\mc{1}{c}{Exp. Time} &\mc{1}{c}{Seeing}  &\mc{1}{c}{$M_{1450}$} \\
\mc{1}{c}{ }      &\mc{1}{c}{$z$}    &\mc{1}{c}{ }   &\mc{1}{c}{Power}     &\mc{1}{c}{(Arcsec)} &\mc{1}{c}{(s)}       &\mc{1}{c}{(Arcsec)} &\mc{1}{c}{ } \\
\tableline
CFHQS\,J0055+0146 & 6.02 & 2007 Dec 14 + 2008 Jan 05 + 2008 Jan 07  & 1300 & 1.0 & \f 5400 & 0.7 & $-24.54$  \\
CFHQS\,J0102-0218 & 5.95 & 2007 Dec 08  & 1300 & 1.0 & \f 1800 & 0.8 & $-24.31$  \\
CFHQS\,J0216-0455 & 6.01 & 2007 Nov 03 + 2007 Nov 11  & 1300 & 1.0 & 14400 & 0.7 & $-22.21$ \\
CFHQS\,J0227-0605 & 6.20 & 2007 Dec 07  & 1300 & 1.0 & \f 5400 & 0.7 &  $-25.03$ \\
CFHQS\,J2318-0246 & 6.05 & 2008 Jun 11  & 1300 & 1.0 & \f 5400 & 0.8 &  $-24.83$ \\
CFHQS\,J2329-0403 & 5.90 & 2007 Dec 10  & 1300 & 1.0 & \f 3600 & 0.7 &  $-24.36$ \\
\hline
\end{tabular}
\tablenotetext{ }{Absolute magnitudes ($M_{1450}$) are calculated using the measured $J$-band magnitudes and assuming a template quasar spectrum. Note that this is slightly different to the method used in W07, which was based on measuring the continuum redward of \lya\ in the spectrum and assuming a spectral index.}
\end{center}
\end{table*}

\subsubsection{Subaru/XMM-Newton Deep Survey (SXDS)}

The SXDS is a deep $BVRi'z'$ survey of $\sim 1.2$ square degrees
carried out at the Subaru Telescope (Furusawa et al. 2008). With
integration times (on the much larger Subaru Telescope) of $\sim$
25\,000 \,s at $i'$ and 13\,000\,s at $z'$, this data reaches
comparable depths to the CFHTLS Deep. Because of its similar depth and
similar $i'z'$ filters to the CFHTLS Deep, the SXDS field is combined
with the Deep to make up the deepest portion of the CFHQS.
CFHQS\,J0216-0455 which is presented in this paper is located in the
SXDS.

\subsection{Quasar selection from imaging}

Candidate quasars are selected from the $i'$ and $z'$ imaging as
objects with colors $i'-z'>2.0$. This is marginally redder than the
selection criterion of $i'-z'>1.7$ used by W07 (but note all four
quasars in W07 have $i'-z'>2.0$). The new criterion will lower quasar
completeness at redshifts $z<6$, but has little impact on the
completeness at higher redshift. The reason for this change in
criterion is due to the large number of M dwarfs observed to have
$1.7<i'-z'<2.0$ due to scattering from the M dwarf locus by
photometric errors. To find all the quasars in this region would be
inefficient use of 8m telescope spectroscopy time.

All candidates with $i'-z'>2.0$ are observed with pointed observations
at $J$-band (see Delorme et al. 2008 for details). $J$-band data are
used to discriminate between high redshift quasars, which are blue in
$z'-J$, and L/T dwarfs, which are red in $z'-J$ (W07; Delorme et
al. 2008). Objects lying in the quasar selection region of the $i'z'J$
color-color diagram are selected as quasar candidates for
spectroscopy. Fig.\,\ref{fig:colcol} shows this diagram for the four
quasars from W07 and the six new quasars in this paper. Also shown is
SDSS\,J2315-0023, which is a $z=6.12$ quasar discovered by Jiang et
al. (2008) in the SDSS Deep Stripe, which we had identified as a
quasar candidate in our CFHQS imaging before it was published by Jiang
et al.  The quasars are mostly located very close to the redshifted
template quasar track.

\subsection{Spectroscopy}
\label{spec}

Optical spectroscopy of candidate quasars was carried out using the
Gemini Multi-Object Spectrograph (GMOS; Hook et al. 2004) at the
Gemini South Telescope. These observations led to the discovery of
six new high redshift quasars.  Positions and photometry for the
quasars are given in Table\,\ref{tab:photom}. $20'' \times 20''$
images centered on the quasars are shown in
Fig.\,\ref{fig:cutouts}. Finding charts in the $z'$-band over a wider
field are presented in the appendix.

GMOS spectroscopy was carried out using the nod-and-shuffle mode. The
details of the spectroscopy observations are given in
Table\,\ref{tab:specobs}. The data processing followed an identical
method to that described in W07.

\begin{figure*}
\resizebox{0.9\textwidth}{!}{\includegraphics{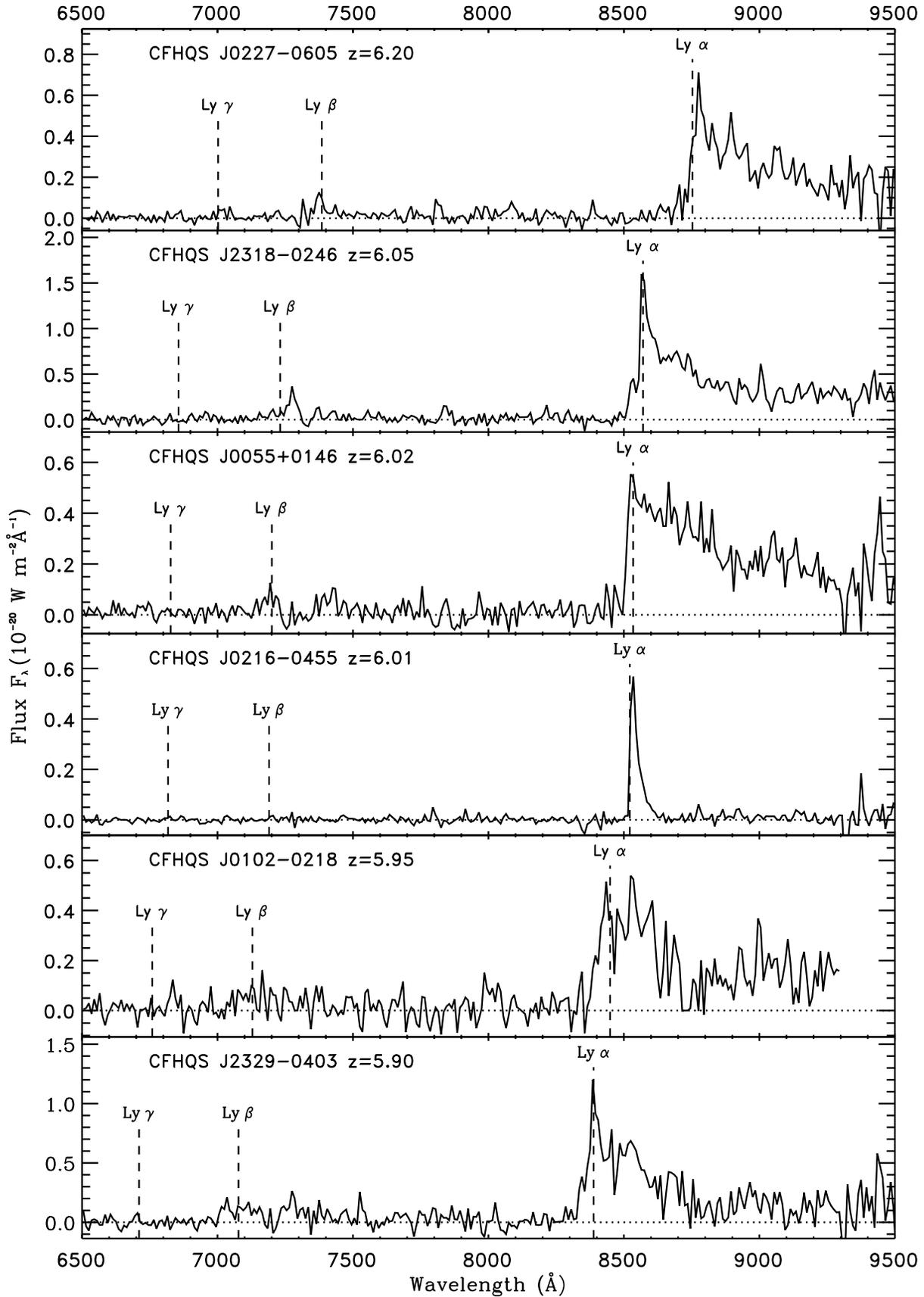}}
\caption{Optical spectra of the six newly discovered quasars. The
expected locations of \lya, \lyb\ and \lyg\ are marked with dashed
lines.  All spectra are binned in 10\,\AA\ pixels.
\label{fig:spec}
}
\end{figure*}

The optical spectra of the six quasars are shown in
Fig.\,\ref{fig:spec}. In all cases there is an unambiguous spectral
break close to the \lya\ emission line, indicating the object is a
$z\sim 6$ quasar. With the lack of other emission lines in these
optical spectra, redshifts for the quasars are estimated using the
\lya\ lines. The systemic redshift is identified at a location close
to the sharp drop on the blue side of broad \lya\ or in some cases a
narrow \lya\ peak, using experience gained from near-IR observations
of previous SDSS and CFHQS $z\sim 6$ quasars. Most of these redshifts
are uncertain by $\sigma_z \approx 0.03$.

\section{Notes on individual quasars}
\label{indiv}

\subsection{CFHQS\,J0227-0605}

This quasar is the first to be found from the CFHTLS Wide. With
$z=6.20$, it is the second highest redshift quasar in the CFHQS so far
and the fifth highest redshift quasar known. It has a fairly strong
\lya\ emission line. There is an apparent dark region of the IGM at
$z\gtsimeq 6$, but higher S/N spectroscopy is required to determine
whether this region does have a high optical depth.

CFHQS\,J0227-0605 has the reddest $z'-J$ color of the CFHQS quasars
discovered so far with $z'-J=1.25$. Partially, this is due to its high
redshift. The simulations of quasar colors as a function of redshift
from Willott et al. (2005) show that $z'-J$ typically starts
increasing at $z>6.1$ due to the dark, IGM-absorbed region of the
spectrum entering the $z'$ filter bandpass. But CFHQS\,J0227-0605 is
0.5 mag redder than the typical quasar at its redshift. This could be
an indication that the quasar continuum is reddened by dust like
several other very high redshift quasars (Maiolino et al. 2004;
McGreer et al. 2006; Venemans et al. 2007; W07). Future near-IR
spectroscopy will test for this. As shown in Fig.\,\ref{fig:colcol},
quasars much redder than CFHQS\,J0227-0605 could fall outside of our
selection box. Most such quasars would still be identified by the
CFBDS which obtains near-IR spectra for most T dwarfs (Albert et
al. in prep.).

\subsection{CFHQS\,J2318-0246}

This quasar was not detected at $i'$ band and has a color limit of
$i'-z'>2.74$. The spectrum shows a typical high redshift quasar at
$z=6.05$ with asymmetric \lya\ line and large continuum break across
the \lya\ line.

\subsection{CFHQS\,J0055+0146}

The colors and spectrum of this $z=6.02$ quasar are typical of
$z\approx 6$ quasars. It is located only $12''$ from a very bright
($i'\approx 14$) star and therefore is well suited to a natural guide
star adaptive optics study of its host galaxy.

\subsection{CFHQS\,J0216-0455}

This object was first identified by McLure et al. (2006) as a
potential $z\approx 6$ Lyman Break galaxy based on optical and near-IR
photometry (their source name: MCD1). However, the fact that it is
unresolved in the $z'$ band ($0.8''$ FWHM) and is relatively bright at
$z'$, suggested that it was also consistent with being a $z\approx 6$
quasar. 

The optical spectrum shows a very strong asymmetric emission
line and weak continuum on the red side of the line only. It is at a
redshift of $z=6.01$. The emission line contributes about 70\% of the
flux detected in the $z'$ band. With an absolute magnitude of
$M_{1450}=-22.21$, this quasar is the least luminous known quasar at
$z>6$ by some margin. Although the luminosity function at $z=6$ is not
well constrained (Shankar \& Mathur 2007), this quasar is likely well
below the break magnitude.

Although the emission line is very narrow by quasar standards, a fit
to the red side of \lya\ suggests an intrinsic symmetric Gaussian
would have FWHM\,$=1600\,{\rm km\,s}^{-1}$ and in addition there is a
residual broader component. These velocities are too great to be caused
by normal galactic processes and indicate an active nucleus. Note that
if CFHQS\,J0216-0455 is accreting at the Eddington limit, then its low
absolute magnitude implies a black hole mass of $2\times
10^7\,M_{\odot}$. Therefore it is not too surprising to find a
relatively narrow emission line in this quasar. However, near-IR
spectroscopy of the \mgii\ line will be required to derive a virial black
hole mass.

\subsection{CFHQS\,J0102-0218}

This $z=5.95$ quasar is not detected at $i'$ band due to the
relatively poor sensitivity of the $i'$ band observation. The color
limit of $i'-z'>2.09$ enabled it to be identified as a quasar
candidate and the $z'-J$ color of $z'-J=0.26$ is typical of $5.5<z<6.1$
quasars. The spectrum shows a typical \lya\ line and a break in the
continuum across the line.

\subsection{CFHQS\,J2329-0403}

This is the lowest redshift CFHQS quasar so far at $z=5.90$. The
spectrum shows a typical broad asymmetric emission line and continuum
break. This quasar should not be confused with CFHQS\,J2329-0301 at
$z=6.42$ presented in W07, which by chance lies only about 1 degree
away.

\section{Multi-wavelength data}
\label{multi}

We have searched the NASA/IPAC Extragalactic Database (NED) at the
locations of the six new quasars and found no coincident objects from
any other survey. Only CFHQS\,J0227-0605 is located in a region
covered by the FIRST radio survey (Becker et al. 1995) and it is not
detected to a flux limit of $\approx 0.5$\,mJy. 

CFHQS\,J0216-0455 is located within the footprint of several deep
multi-wavelength surveys. It is not detected at 1.4\,GHz by the
100$\mu$Jy VLA survey of Simpson et al. (2006). It is also not
detected in the deep XMM-Newton survey of Ueda et al. (2008) which
reaches a typical flux limit of $2\times 10^{-15}\,{\rm
  ergs\,cm^{-2}\,s^{-1}}$ in the $0.5-4.5$ keV band corresponding to
$L_{0.5-4.5}<5 \times 10^{44}\,{\rm ergs\,s^{-1}}$. This X-ray
non-detection is not surprising considering that CFHQS\,J0216-0455 is
a very faint quasar. The typical X-ray luminosity for a quasar with
$M_{1450} =-22.41$ is only $L_{0.5-4.5}= 10^{44}\,{\rm ergs\,s^{-1}}$,
adopting the luminosity-dependent X-ray-optical relationship of
Vignali et al. (2003). CFHQS\,J0216-0455 is detected by the deep {\it
  Spitzer} UDS survey with AB magnitudes of $24.10 \pm 0.25$ at $3.6
\mu{\rm m}$ and $23.18 \pm 0.15$ at $4.5 \mu{\rm m}$. The color
$J-3.6=0.1$ is typical of quasars at this redshift (Jiang et
al. 2006). The $3.6-4.5$ color is redder than all the more luminous
quasars studied by Jiang et al. (2006) and may be due to a very high
equivalent width H$\alpha$ line in CFHQS\,J0216-0455.

\section{Conclusions}

We have presented the discovery of six high-redshift quasars from the
CFHQS. These quasars span a wide range in luminosity including the
lowest luminosity quasar known at $z>6$: CFHQS\,J0216-0455. Future
near-infrared spectroscopy will test whether lower luminosity quasars
are powered by lower mass black holes accreting at the Eddington limit.

The CFHQS has now discovered a total of 10 quasars at redshifts $z \ge
5.9$. Work is underway to complete followup of quasar candidates and
determine the completeness of the sample. A future publication will
use these quasars to determine the bright end slope of the $z=6$
quasar luminosity function.

The CFHQS continues to discover quasars at redshifts $z>6$ which allow
us to probe the ionization state of the intergalactic medium. The
discovery spectra presented here are not sensitive enough for these
faint quasars to allow an investigation of this issue. We are carrying
out higher S/N and higher resolution spectroscopy to probe the
intergalactic medium along these lines of sight.

\acknowledgments

Based on observations obtained with MegaPrime/MegaCam, a joint project
of CFHT and CEA/DAPNIA, at the Canada-France-Hawaii Telescope (CFHT)
which is operated by the National Research Council (NRC) of Canada,
the Institut National des Sciences de l'Univers of the Centre National
de la Recherche Scientifique (CNRS) of France, and the University of
Hawaii. This work is based in part on data products produced at
TERAPIX and the Canadian Astronomy Data Centre as part of the
Canada-France-Hawaii Telescope Legacy Survey, a collaborative project
of NRC and CNRS. Based on observations obtained at the Gemini
Observatory, which is operated by the Association of Universities for
Research in Astronomy, Inc., under a cooperative agreement with the
NSF on behalf of the Gemini partnership: the National Science
Foundation (United States), the Particle Physics and Astronomy
Research Council (United Kingdom), the National Research Council
(Canada), CONICYT (Chile), the Australian Research Council
(Australia), CNPq (Brazil) and CONICET (Argentina). This paper uses
data from Gemini programs GS-2007A-Q-24 and GS-2007B-Q-15.  Based on
observations made with the ESO New Technology Telescope at the La
Silla Observatory. This work is based in part on data obtained as part
of the UKIRT Infrared Deep Sky Survey. This research has made use of
the NASA/IPAC Extragalactic Database (NED) which is operated by the
Jet Propulsion Laboratory, California Institute of Technology, under
contract with the National Aeronautics and Space
Administration. Thanks to Howard Yee and the rest of the RCS2 team for
sharing their data and to the queue observers at CFHT and Gemini
who obtained data for this project. Thanks to the anonymous referee for comments to improve the manuscript.



\appendix

\section{Finding charts}

\begin{figure*}[b]
\hspace{1.1cm}
\resizebox{0.88\textwidth}{!}{\includegraphics{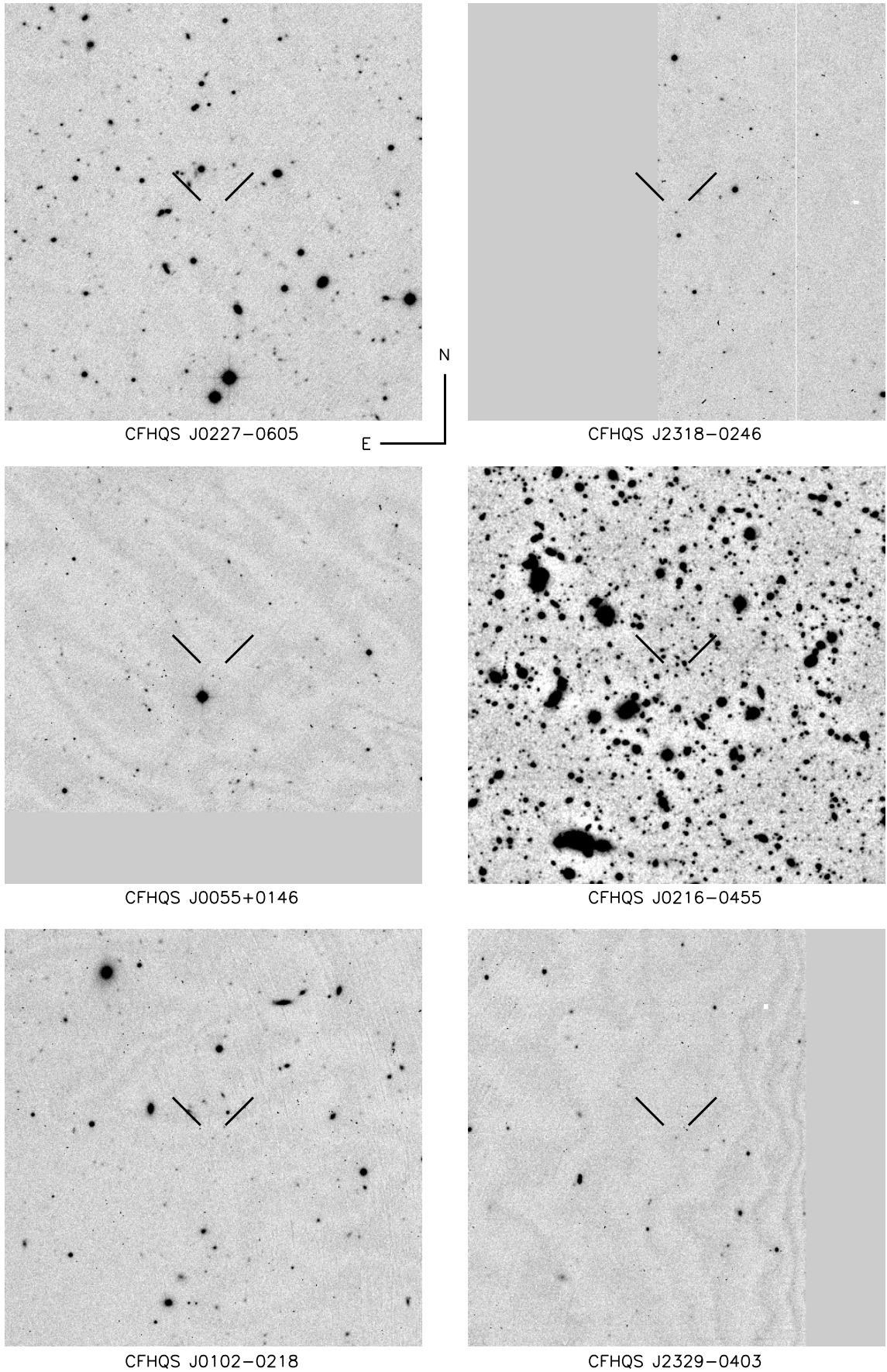}}
\caption{ $z'$-band finding charts for the CFHQS quasars.
\label{fig:finders}
}
\end{figure*}

Fig.\,\ref{fig:finders} presents $3' \times 3'$ finding charts for the
CFHQS quasars. All images are centered on the quasars and have the same
orientation on the sky. These are the CFHT MegaCam or Subaru
SuprimeCam $z'$-band images in which the quasars were first
identified. MegaCam has gaps between the CCDs and these data were not
dithered so the gaps remain and are evident in three images where the
quasars lie close to the edge of a CCD. For CFHQS\,J2318-0246,
CFHQS\,J0055+0146, CFHQS\,J0102-0218 and CFHQS\,J2329-0403, the
Megacam data are a single exposure leading to many cosmic rays in the
final images.

\end{document}